\documentclass[a4paper,11pt]{article}
\usepackage{amsmath}

\numberwithin{equation}{section}

\newcommand{\kakko}[1]{\ensuremath{\left(#1\right)}}
\newcommand{\Kakko}[1]{\ensuremath{\left\{#1\right\}}}

\newcommand{\absolute}[1]{\ensuremath{\left|#1\right|}}
\newcommand{\ket}[1]{\ensuremath{\left|#1\right\rangle}}
\newcommand{\Set}[2]{\Kakko{#1\left|#2\vphantom{#1}\right.}}
\newcommand{\commentout}[1]{}

\renewcommand{\bar}[1]{{\overline{#1}}}
\newcommand{\oa}[1]{{\overline{\alpha#1}}}
\newcommand{\ob}[1]{{\overline{\beta#1}}}
\newcommand{\oba}[1]{{\overline{\beta#1-\alpha#1}}}
\newcommand{\oma}[1]{{\overline{-\alpha#1}}}

\newenvironment{Cases}{\left\{\begin{aligned}}{\end{aligned}\right.}

\begin{document}

\vspace*{10mm}
\begin{center}
	{\LARGE Modular invariance of bosonic string on orbifolds} \\[5mm]

	Susumu Fujita\footnote{E-mail: sfujita@graduate.chiba-u.jp} \\[5mm]

	{\it Graduate School of Science and Technology, Chiba University,
	Chiba 263-8522, Japan}
\end{center}
\vspace*{10mm}

\begin{abstract}
I construct a complete 1-loop partition function of a bosonic closed string
on orbifolds.
Furthermore, I derive sufficient conditions for the modular invariance of
the partition function.
\end{abstract}

\section{Introduction}

Orbifold compactification models of the heterotic string \cite{orbifold1} are
candidates for unified theories.
In the string theory, modular invariance is necessary for consistency.
But modular invariance of orbifold models is non-trivial.
A modular invariance condition of orbifold models is naively the level
matching \cite{orbifold2,level_matching}.
In contrast, a free fermion model was built as another possibility
\cite{fermionic}.
Although the free fermion model is given by fermionizing all internal
coordinates, the fermionization on orbifolds is restricted on
the $\mathbf{Z}_2$ orbifold.
Therefore we should consider the internal coordinates as bosonic variables
so as to generalize the model to include general orbifolds.
A 1-loop partition function of a bosonic closed string on orbifolds was
constructed in the ref.\cite{bosonic}.
But I think that this partition function is not complete in the points
explained in the text.
In this paper I construct a complete 1-loop partition function of a bosonic
closed string on orbifolds.
Furthermore, I derive sufficient conditions for the modular invariance of
the partition function.

\section{Preliminaries}

In this section I set up the framework for this paper.
I consider a heterotic string compactified to four space-time dimensions.
In the light-cone gauge, there are the following world-sheet degrees of
freedom:
eight transverse bosons $X^i(z,\bar{z})$,
eight right-moving transverse fermions $\tilde{\psi}^i(\bar{z})$ where
$i=2,3,\dots,9$,
and sixteen left-moving bosons $X_L^I(z)$ where $I=10,11,\dots,25$.
Here the world-sheet coordinates are $z=e^{-i(\sigma^1+i\sigma^2)}$ and
$\bar{z}=e^{i(\sigma^1-i\sigma^2)}$.
In this paper I particularly pay attention to the bosons $X^n$ where
$n=4,5,\dots,9$ (corresponding to six internal coordinates).

An orbifold is obtained from flat space by the following identification under
a discrete group $G$.
An element $g$ of $G$ acts on the coordinates as a rotation $\theta$ and
a translation $2\pi R$,
\begin{equation}
	g:X^n \to \theta^{nm}X^m+2\pi R^n,
\end{equation}
where $m,n=4,5,\dots,9$.
Let $N$ be the smallest integer such as $\theta^N=1$,
then this orbifold is called a $\mathbf{Z}_N$ orbifold.
It is convenient to change the basis of the coordinates so that the rotation
$\theta$ becomes a diagonal matrix.
Then the real bosons $X^n$ become complex bosons $\phi^a$ where $a=1,2,3$.
In this basis, the action of $g$ becomes
\begin{equation}
	g:\phi^a\to e^{2\pi i\oa{v^a}}\phi^a+2\pi\ell^a,
		\label{identification}
\end{equation}
where $\alpha$ is an integer, $v^a$ is a multiple of $1/N$ in the
$\mathbf{Z}_N$ orbifold, and $\ell$ is an element of a 3-dimensional
complex lattice $\Lambda$.
The lattice $\Lambda$ must be invariant under the rotation
$\mathrm{diag}(e^{2\pi iv^1},\\ e^{2\pi iv^2},e^{2\pi iv^3})$
so that the orbifold is well-defined.
$\oa{v^a}$ is defined by $\oa{v^a}=\alpha v^a \pmod 1$ and $0\le\oa{v^a}<1$.
Then the elements of the group $G$ are in one-to-one correspondence with
the parameters $\oa{v}$ and $\ell$, which are denoted by $g(\oa{v},\ell)$.
A representation of $g(\oa{v},\ell)$ is defined by
\begin{equation}
	g^{-1}(\oa{v},\ell)\phi^a g(\oa{v},\ell)
		=e^{2\pi i\oa{v^a}}\phi^a+2\pi\ell^a,
\end{equation}
and $g(\oa{v},\ell)$ is a unitary operator.

\section{Partition function}

In this section I construct the 1-loop partition function of the bosonic
field $\phi$ on the orbifold.
The heterotic string theory contains only the closed string.
The 1-loop world-sheet of a closed string is a torus.
The torus is described by the identifications
$z\cong e^{2\pi i}z$ and $z\cong e^{2\pi i\tau}z$, where $\tau$ is a complex
number.
Therefore the field $\phi^a$ satisfies the periodic boundary conditions
$\phi^a(e^{2\pi i}z,e^{-2\pi i}\bar{z})=\phi^a(z,\bar{z})$ and
$\phi^a(e^{2\pi i\tau}z,e^{-2\pi i\bar{\tau}}\bar{z})=\phi^a(z,\bar{z})$.
On the orbifold, because of the identification under the group $G$,
the following boundary conditions are allowed,
\begin{subequations}
\begin{align}
	\phi^a(e^{2\pi i}z,e^{-2\pi i}\bar{z})
		& =g^{-1}(\oa{v},\ell) \phi^a(z,\bar{z}) g(\oa{v},\ell),
		\label{boundary_a} \\
	\phi^a(e^{2\pi i\tau}z,e^{-2\pi i\bar{\tau}}\bar{z})
		& =g^{-1}(\ob{v},\ell') \phi^a(z,\bar{z}) g(\ob{v},\ell'),
		\label{boundary_b}
\end{align}
\end{subequations}
where $\alpha$ and $\beta$ are integers and $\ell,\ell'\in\Lambda$.
Then the partition function is given by
\begin{equation}
	Z(\tau)=\frac{1}{N} \sum_{\alpha=0}^{N-1} \sum_{\beta=0}^{N-1}
		Z^\oa{v}_\ob{v}(\tau),
		\label{def_Z}
\end{equation}
with
\begin{equation}
	Z^\oa{v}_\ob{v}(\tau)=\sum_\ell \sum_{\ell'} \mathrm{Tr}
		\Kakko{q^{H_{\oa{v},\ell}} \bar{q}^{\tilde{H}_{\oa{v},\ell}}
		g(\ob{v},\ell')},
		\label{Z_ab}
\end{equation}
where $q=e^{2\pi i\tau}$, and $H_{\oa{v},\ell}$ and $\tilde{H}_{\oa{v},\ell}$
are left- and right-moving Hamiltonians respectively.

Let $\phi_{\oa{v},\ell}^a$ be a field which satisfies the boundary condition
\eqref{boundary_a}.
Then I obtain
\begin{equation}
	\phi_{\oa{v},\ell}^a(e^{2\pi i}z,e^{-2\pi i}\bar{z})
		=e^{2\pi i\oa{v^a}} \phi_{\oa{v},\ell}^a(z,\bar{z})+2\pi\ell^a.
		\label{condition_phi}
\end{equation}
In case of $\oa{v^a}=0$, this field has the mode expansion
\begin{equation}
	\phi_{\oa{v},\ell}^a(z,\bar{z})
		=\chi^a-i\rho_L^a\log z-i\rho_R^a\log \bar{z}
		+i\sum_{n=1}^\infty\frac{1}{n}(\beta_n^a z^{-n}
		+\gamma_n^{\dagger a} z^n+\tilde{\beta}_n^a \bar{z}^{-n}
		+\tilde{\gamma}_n^{\dagger a} \bar{z}^n),
		\label{mode_expansion1}
\end{equation}
where I defined the left- and right-moving momenta respectively by
\begin{equation}
	\rho_L^a=\rho^a+\frac{\ell^a}{2}, \quad \rho_R^a=\rho^a-\frac{\ell^a}{2},
		\label{rhoL_rhoR}
\end{equation}
$\chi$ is the center of mass position,
$\gamma_n^\dagger$ and $\tilde{\gamma}_n^\dagger$ are the creation operators,
and $\beta_n$ and $\tilde{\beta}_n$ are the annihilation operators.
The commutation relations of these operators are
\begin{subequations}
\begin{gather}
	[\chi^a,\rho^{\dagger b}]=[\chi_L^a,\rho_L^{\dagger b}]
		=[\chi_R^a,\rho_R^{\dagger b}]=i\delta^{ab},
		\label{commutator_xp} \\
	[\beta^a_m,\beta^{\dagger b}_n]
		=[\gamma^a_m,\gamma^{\dagger b}_n]
		=[\tilde{\beta}^a_m,\tilde{\beta}^{\dagger b}_n]
		=[\tilde{\gamma}^a_m,\tilde{\gamma}^{\dagger b}_n]
		=m\delta_{mn}\delta^{ab},
		\label{commutator_bb}
\end{gather}
\end{subequations}
where $\chi_L^a$ and $\chi_R^a$ are the left- and right-moving parts of
$\chi^a$, respectively.
In case of $\oa{v^a}\ne0$, on the other hand, the field
$\phi_{\oa{v},\ell}^a$ has the mode expansion
\begin{equation}
	\begin{split}
		\phi_{\oa{v},\ell}^a(z,\bar{z})=\chi^a+i\sum_{n=1}^\infty
			& \left(\frac{\beta_{n-\oa{v}}^a}{n-\oa{v^a}}z^{-n+\oa{v^a}}
			+\frac{\gamma_{n+\oa{v}-1}^{\dagger a}}{n+\oa{v^a}-1}
			z^{n+\oa{v^a}-1} \right. \\
		& \quad \left. +\frac{\tilde{\beta}_{n+\oa{v}-1}^a}{n+\oa{v^a}-1}
			\bar{z}^{-n-\oa{v^a}+1}
			+\frac{\tilde{\gamma}_{n-\oa{v}}^{\dagger a}}
			{n-\oa{v^a}}\bar{z}^{n-\oa{v^a}} \right).
	\end{split}
	\label{mode_expansion2}
\end{equation}
The commutation relations of the creation and annihilation operators are
similar to \eqref{commutator_bb}.
Then, to satisfy the boundary condition \eqref{condition_phi},
$\chi^a$ must be a fixed point which satisfies
\begin{equation}
	e^{2\pi i\oa{v^a}}\chi^a+2\pi\ell^a=\chi^a.
		\label{fixed_point_a}
\end{equation}

Now I decompose the left-moving Hamiltonian $H_{\oa{v},\ell}$ into
$H_{\rho;\,\oa{v},\ell}$ and $H_{N;\,\oa{v}}$,
where $H_{\rho;\,\oa{v},\ell}$ depends on the momentum $\rho^{}_L$.
Let the right-moving Hamiltonian $\tilde{H}_{\oa{v},\ell}$ be similarly
divided.
\begin{equation}
	H_{\oa{v},\ell}=H_{\rho;\,\oa{v},\ell}+H_{N;\,\oa{v}}, \quad
		\tilde{H}_{\oa{v},\ell}=\tilde{H}_{\rho;\,\oa{v},\ell}
		+\tilde{H}_{N;\,\oa{v}}.
\end{equation}
Here
\begin{subequations}
\begin{align}
	H_{\rho;\,\oa{v},\ell} & =\rho_L^\dagger\circ\rho_L
		=(\rho^\dagger+\bar{\ell}/2)\circ(\rho+\ell/2), \\
	\tilde{H}_{\rho;\,\oa{v},\ell} & =\rho_R^\dagger\circ\rho_R
		=(\rho^\dagger-\bar{\ell}/2)\circ(\rho-\ell/2), \\
	\intertext{(the symbol $\circ$ is defined by
		$x\circ y=\sum_{a=1}^3 \delta_{\oa{v^a},0} x^a y^a$)}
	H_{N;\,\oa{v}}
		& =\sum_{n=1}^\infty \Kakko{(n-\oa{v})\cdot N_{n-\oa{v}}
		+(n+\oa{v}-1)\cdot N'_{n+\oa{v}-1}}
		+\frac{1}{2}\oa{v}\cdot(1-\oa{v})-\frac{3}{12}, \\
	\tilde{H}_{N;\,\oa{v}}
		& =\sum_{n=1}^\infty \Kakko{(n+\oa{v}-1)\cdot\tilde{N}_{n+\oa{v}-1}
		+(n-\oa{v})\cdot\tilde{N}'_{n-\oa{v}}}
		+\frac{1}{2}\oa{v}\cdot(1-\oa{v})-\frac{3}{12}.
\end{align}
\end{subequations}
$N_r^a,N_r^{\prime a},\tilde{N}_r^a$ and $\tilde{N}_r^{\prime a}$ are
the occupation numbers defined as
$N^a_r=\beta^{\dagger a}_r\beta^a_r/r^a$,
$N^{\prime a}_r=\gamma^{\dagger a}_r\gamma^a_r/r^a$,
$\tilde{N}^a_r=\tilde{\beta}^{\dagger a}_r\tilde{\beta}^a_r/r^a$ and
$\tilde{N}^{\prime a}_r=\tilde{\gamma}^{\dagger a}_r\tilde{\gamma}^a_r/r^a$,
but $N_0^{\prime a}=\tilde{N}_0^a=0$.

Now an element $g^{}_{\oa{v},\ell}(\ob{v},\ell')$ of the group $G$ is defined
by
\begin{equation}
	g_{\oa{v},\ell}^{-1}(\ob{v},\ell') \phi_{\oa{v},\ell}^a(z,\bar{z})
		g^{}_{\oa{v},\ell}(\ob{v},\ell')
		=e^{2\pi i\ob{v^a}}\phi_{\oa{v},\ell}^a(z,\bar{z})
		+2\pi\ell^{\prime a}.
		\label{def_g}
\end{equation}
up to an arbitrary constant $C^{\oa{v},\ell}_{\ob{v},\ell'}$.
This constant is a phase factor
$\kakko{\absolute{C^{\oa{v},\ell}_{\ob{v},\ell'}}^2=1}$, because
$g^{}_{\oa{v},\ell}(\ob{v},\ell')$ is a unitary operator.
In ref.\cite{bosonic}, this constant was chosen to be independent of $\ell$
and $\ell'$.
But $\ell$ and $\ell'$ dependence of this constant is necessary for
convergence of the partition function \eqref{Z_ab} summed over all $\ell$ and
$\ell'$.
In this paper, therefore, I consider $\ell$ and $\ell'$ dependence of this
constant.
Now I divide $g^{}_{\oa{v},\ell}(\ob{v},\ell')$ except this constant
into a factor $h_{\rho;\,\oa{v}}(\ob{v},\ell')$ which depends the momenta
and the other factor $h_{N;\,\oa{v}}(\ob{v})$,
\begin{equation}
	g^{}_{\oa{v},\ell}(\ob{v},\ell')=C^{\oa{v},\ell}_{\ob{v},\ell'}
		h_{\rho;\,\oa{v}}(\ob{v},\ell')h_{N;\,\oa{v}}(\ob{v}),
\end{equation}
where
\begin{subequations}
\begin{align}
	h_{\rho;\,\oa{v}}(\ob{v},\ell')
		& =\exp\Kakko{-2\pi i(\bar{\ell'}\circ\rho+\ell'\circ\rho^\dagger)}
		\exp\Kakko{2\pi i\ob{v}\circ(K_L+K_R)}, \\
	h_{N;\,\oa{v}}(\ob{v})
		& =\exp\Kakko{2\pi i\ob{v}\cdot(J_\oa{v}+\tilde{J}_\oa{v})},
\end{align}
\vspace{-8mm}
\begin{alignat}{2}
	K_L^a & =-i\kakko{\chi_L^a \rho_L^{\dagger a}
		-\rho_L^a \chi_L^{\dagger a}}, & \quad
		K_R^a & =-i\kakko{\chi_R^a \rho_R^{\dagger a}
		-\rho_R^a \chi_R^{\dagger a}}, \\
	J^a_\oa{v} & =\sum_{n=1}^\infty\kakko{N^a_{n-\oa{v}}
		-N^{\prime a}_{n+\oa{v}-1}},  & \quad
		\tilde{J}^a_\oa{v} & =\sum_{n=1}^\infty\kakko{\tilde{N}^a_{n+\oa{v}-1}
		-\tilde{N}^{\prime a}_{n-\oa{v}}}.
\end{alignat}
\end{subequations}
In case of $\oa{v^a}\ne0$, $\chi^a$ must be a fixed point which satisfies
\begin{equation}
	e^{2\pi i\ob{v^a}}\chi^a+2\pi\ell^{\prime a}=\chi^a,
		\label{fixed_point_b}
\end{equation}
so that the field \eqref{mode_expansion2} is transformed according to
\eqref{def_g},
because there is no noncommutative operator to $\chi^a$.

Thus, the partition function \eqref{Z_ab} is given by product of two functions
${Z_\rho}^\oa{v}_\ob{v}(\tau)$ and ${Z_N}^\oa{v}_\ob{v}(\tau)$,
\begin{equation}
	Z^\oa{v}_\ob{v}(\tau)={Z_\rho}^\oa{v}_\ob{v}(\tau)
		{Z_N}^\oa{v}_\ob{v}(\tau),
\end{equation}
with
\begin{subequations}
\begin{align}
	{Z_\rho}^\oa{v}_\ob{v}(\tau)
		& =\sum_\ell \sum_{\ell'} C^{\oa{v},\ell}_{\ob{v},\ell'} \mathrm{Tr}
		\Kakko{q^{H_{\rho;\,\oa{v},\ell}}
		\bar{q}^{\tilde{H}_{\rho;\,\oa{v},\ell}}
		h_{\rho;\,\oa{v}}(\ob{v},\ell')},
		\label{def_Z_rho} \\
	{Z_N}^\oa{v}_\ob{v}(\tau)
		& =\mathrm{Tr} \Kakko{q^{H_{N\,\oa{v}}}
		\bar{q}^{\tilde{H}_{N\,\oa{v}}} h_{N\,\oa{v}}(\ob{v})}.
		\label{def_Z_N}
\end{align}
\end{subequations}

First, I calculate the function ${Z_\rho}^\oa{v}_\ob{v}(\tau)$.
I consider common eigenstates of the operators $H_{\rho;\,\oa{v},\ell}$,
$\tilde{H}_{\rho;\,\oa{v},\ell}$ and $h_{\rho;\,\oa{v}}(\ob{v},\ell')$
so as to calculate the function ${Z_\rho}^\oa{v}_\ob{v}(\tau)$.
In ref.\cite{bosonic}, these eigenstates are not discussed.
But, if these eigenstates are not discussed then the function
${Z_\rho}^\oa{v}_\ob{v}(\tau)$ cannot be exactly derived.
In this paper, therefore, I derive these eigenstates and their eigenvalues.
Let $\ket{\rho^{}_L,\rho^{}_R}_\oa{v}$ be a common eigenstate of both
operators $\hat{\rho}^{}_L$ and $\hat{\rho}^{}_R$.
This state is an eigenstate of both $H_{\rho;\,\oa{v},\ell}$ and
$\tilde{H}_{\rho;\,\oa{v},\ell}$.
But this state is not an eigenstate of $h_{\rho;\,\oa{v}}(\ob{v},\ell')$,
because the operator $e^{2\pi i\ob{v}\circ(K_L+K_R)}$ in
$h_{\rho;\,\oa{v}}(\ob{v},\ell')$ rotate the momenta $\rho^{}_L$ and
$\rho^{}_R$ of this state,
\begin{subequations}
\begin{align}
	\hat{\rho}_L^a \kakko{e^{2\pi i\ob{v}\circ(K_L+K_R)}
		\ket{\rho^{}_L,\rho^{}_R}_\oa{v}}
		& =e^{2\pi i\ob{v^a}} \rho_L^a
		\kakko{e^{2\pi i\ob{v}\circ(K_L+K_R)}
		\ket{\rho^{}_L,\rho^{}_R}_\oa{v}}, \label{e^bv_rhoL} \\
	\hat{\rho}_R^a \kakko{e^{2\pi i\ob{v}\circ(K_L+K_R)}
		\ket{\rho^{}_L,\rho^{}_R}_\oa{v}}
		& =e^{2\pi i\ob{v^a}} \rho_R^a
		\kakko{e^{2\pi i\ob{v}\circ(K_L+K_R)}
		\ket{\rho^{}_L,\rho^{}_R}_\oa{v}}. \label{e^bv_rhoR}
\end{align}
\end{subequations}
Therefore I consider linear combinations of the states,
\begin{equation}
	\ket{\bar{\rho^{}_L}\circ\rho^{}_L,\bar{\rho^{}_R}\circ\rho^{}_R,k}
		^\oa{v}_{\ob{v},\ell'}
		=\frac{1}{\sqrt{m}} \sum_{n=0}^{m-1} \exp\kakko{2\pi i\lambda_n^k}
		\exp\Kakko{2\pi i\bar{n\beta v}\circ(K_L+K_R)}
		\ket{\rho^{}_L,\rho^{}_R}_\oa{v},
		\label{eigenstate}
\end{equation}
where $m$ is the smallest positive integer such that
$\bar{m\beta v^a}=0$ for $\forall a \; (\oa{v^a}=0, \; (\rho_L^a\ne0
\text{ or } \rho_R^a\ne0))$, and $k=0,1,\dots,m-1$.
Let $\lambda_0^k$ be zero.
In addition, let $\lambda_n^k$ satisfy
\begin{equation}
	\frac{1}{m}\sum_{n=0}^{m-1} \exp\kakko{-2\pi i\lambda_n^k}
		\exp\kakko{2\pi i\lambda_n^{k'}}=\delta^{kk'},
		\label{orthonormal}
\end{equation}
so that these states \eqref{eigenstate} belong to the orthonormal system.
These states are eigenstates of both
$H_{\rho;\,\oa{v},\ell}=\rho_L^\dagger\circ\rho^{}_L$ and
$\tilde{H}_{\rho;\,\oa{v},\ell}=\rho_R^\dagger\circ\rho^{}_R$ because of
\eqref{e^bv_rhoL} and \eqref{e^bv_rhoR}.
The operation of $h_{\rho;\,\oa{v}}(\ob{v},\ell')$ to these states becomes
\begin{equation}
	\begin{split}
		& h_{\rho;\,\oa{v}}(\ob{v},\ell')
			\ket{\bar{\rho^{}_L}\circ\rho^{}_L,\bar{\rho^{}_R}\circ\rho^{}_R,
			k}^\oa{v}_{\ob{v},\ell'} \\
		& =\frac{1}{\sqrt{m}} \sum_{n=0}^{m-1} \exp\kakko{2\pi i\lambda_n^k}
			\exp\Kakko{-2\pi i(\bar{\ell'}\circ\rho+\ell'\circ\rho^\dagger)}
			\exp\Kakko{2\pi i\ob{v}\circ(K_L+K_R)} \\
		& \qquad\qquad\quad \times
			\exp\Kakko{2\pi i\bar{n\beta v}\circ(K_L+K_R)}
			\ket{\rho^{}_L,\rho^{}_R}_\oa{v} \\
		& =\frac{1}{\sqrt{m}} \sum_{n=0}^{m-1} \exp\kakko{2\pi i\lambda_n^k}
			\exp\Kakko{-2\pi i\kakko{\bar{\ell'}\circ
			e^{2\pi i\bar{(n+1)\beta v}}\rho+\text{c.c.}}} \\
		& \qquad\qquad\quad \times
			\exp\Kakko{2\pi i\bar{(n+1)\beta v}\circ(K_L+K_R)}
			\ket{\rho^{}_L,\rho^{}_R}_\oa{v},
	\end{split}
\end{equation}
where $\lambda_n^k$ are
\begin{equation}
	\lambda_n^k=n\lambda'_k
		-\Kakko{\bar{\ell'}\circ\kakko{e^{2\pi i\bar{\beta v}}
		+e^{2\pi i\bar{2\beta v}}+\cdots+e^{2\pi i\bar{n\beta v}}}\rho
		+\text{c.c.}} \mod 1,
		\label{lambda_n^k}
\end{equation}
so that these states are eigenstates of $h_{\rho;\,\oa{v}}(\ob{v},\ell')$.
Then the eigenvalue of $h_{\rho;\,\oa{v}}(\ob{v},\ell')$ is
$\exp\kakko{-2\pi i\lambda'_k}$.
If $\rho^a\ne0$ then $\rho_L^a\ne0$ or $\rho_R^a\ne0$ because of the relations
\eqref{rhoL_rhoR}.
Therefore, if $\rho^a\ne0$ then $\bar{m\beta v^a}=0$ owing to the definition
of $m$.
Then, if $\ob{v^a}\ne0$ then the summation
$e^{2\pi i\bar{\beta v}}+e^{2\pi i\bar{2\beta v}}+\cdots
+e^{2\pi i\bar{m\beta v}}$ is zero.
On the other hand, if $\ob{v^a}=0$ then this summation is $m$.
Thus $\lambda_m^k$ are
\begin{equation}
	\lambda_m^k=m\lambda'_k-m(\bar{\ell'}\bullet\rho+\ell'\bullet\bar{\rho})
		\mod 1,
\end{equation}
where the symbol $\bullet$ is defined by
$x\bullet y=\sum_{a=1}^3 \delta_{\oa{v^a},0} \delta_{\ob{v^a},0} x^a y^a$.
$\lambda_m^k$ must be equal to $\lambda_0^k$ modulo 1 ($\lambda_0^k=0$).
Hence $\lambda'_k$ are
\begin{equation}
	\lambda'_k=\bar{\ell'}\bullet\rho+\ell'\bullet\bar{\rho}+\frac{k}{m},
\end{equation}
so that $\lambda_n^k$ satisfy the orthonormal condition \eqref{orthonormal}.

In the function ${Z_\rho}^\oa{v}_\ob{v}(\tau)$,
the parameter $k$ is included in the eigenvalue of only
$h_{\rho;\,\oa{v}}(\ob{v},\ell')$.
Therefore, in the trace in ${Z_\rho}^\oa{v}_\ob{v}(\tau)$,
the eigenvalues of only $h_{\rho;\,\oa{v}}(\ob{v},\ell')$ are summed over $k$,
\begin{equation}
	\sum_{k=0}^{m-1} \exp\Kakko{-2\pi i\kakko{\bar{\ell'}\bullet\rho
		+\ell'\bullet\bar{\rho}+\frac{k}{m}}}=
		\begin{cases}
			\exp\Kakko{-2\pi i\kakko{\bar{\ell'}\bullet\rho
				+\ell'\bullet\bar{\rho}}} & (m=1) \\
			0 & (m>1)
		\end{cases}.
\end{equation}
Hence only the states of $m=1$ contribute to ${Z_\rho}^\oa{v}_\ob{v}(\tau)$.
In the states of $m=1$, $\rho_L^a=\rho_R^a=0$ for $\forall a \;
(\oa{v^a}=0,\,\ob{v^a}\ne0)$ because of the definition of $m$.
If $\rho_L^a=\rho_R^a=0$ then $\rho^a=\ell^a=0$ owing to the relations
\eqref{rhoL_rhoR}.
Therefore only the states of $\rho^a=\ell^a=0$ for $\forall a \;
(\oa{v^a}=0,\,\ob{v^a}\ne0)$ contributes to ${Z_\rho}^\oa{v}_\ob{v}(\tau)$.
Since $\rho$ is the vector in $\oa{v^a}=0$ space,
only $\rho$ in $\oa{v^a}=\ob{v^a}=0$ space contributes to
${Z_\rho}^\oa{v}_\ob{v}(\tau)$.
I write the momentum in $\oa{v^a}=\ob{v^a}=0$ space as $\rho^{}_0$.
In addition, the set of such $\ell$ as contribute to
${Z_\rho}^\oa{v}_\ob{v}(\tau)$ is
\begin{equation}
	{\Lambda_\ell}^\oa{v}_\ob{v}
		=\Set{\ell}{\ell\in\Lambda, \; \forall a \;
		(\oa{v^a}=0, \; \ob{v^a}\ne0), \; \ell^a=0}.
		\label{Lambda_la}
\end{equation}
In case of $\oa{v^a}\ne0$, if $\ob{v^a}=0$ then $\ell^{\prime a}=0$
because of the fixed point condition \eqref{fixed_point_b}.
Therefore the set of such $\ell'$ as contribute to
${Z_\rho}^\oa{v}_\ob{v}(\tau)$ is
\begin{equation}
	{\Lambda_{\ell'}}^\oa{v}_\ob{v}
		=\Set{\ell'}{\ell'\in\Lambda, \; \forall a \;
		(\oa{v^a}\ne0, \; \ob{v^a}=0), \; \ell^{\prime a}=0}.
		\label{Lambda_lb}
\end{equation}

But all elements of ${\Lambda_\ell}^\oa{v}_\ob{v}$ and
${\Lambda_{\ell'}}^\oa{v}_\ob{v}$ do not necessarily contribute to
${Z_\rho}^\oa{v}_\ob{v}(\tau)$, because $\chi$ satisfied to the both
conditions \eqref{fixed_point_a} and \eqref{fixed_point_b} must exist.
In ref.\cite{bosonic}, it is not considered that $\chi$ must satisfy the
both conditions \eqref{fixed_point_a} and \eqref{fixed_point_b}.
But, if it is not considered then the function
${Z_\rho}^\oa{v}_\ob{v}(\tau)$ cannot be exactly derived.
In this paper, therefore, I introduce the factor
\begin{equation}
	\xi^{\oa{v},\ell}_{\ob{v},\ell'}=
		\begin{Cases}
			& 1 && \kakko{\exists\chi, \; \forall a \;
				(\oa{v^a}\ne0, \; \ob{v^a}\ne0), \;
				e^{2\pi i\oa{v^a}}\chi^a+2\pi\ell^a
				=e^{2\pi i\ob{v^a}}\chi^a+2\pi\ell^{\prime a}=\chi^a} \\
			& 0 && (\text{otherwise})
		\end{Cases}.
		\label{def_xi}
\end{equation}
In addition, $H_{\rho;\,\oa{v},\ell}=\bar{(\rho^{}_0+\ell/2)}\bullet
(\rho^{}_0+\ell/2)$ and $\tilde{H}_{\rho;\,\oa{v},\ell}
=\bar{(\rho^{}_0-\ell/2)}\bullet(\rho^{}_0-\ell/2)$, because
$\oa{v^a}=0, \, \ob{v^a}\ne0$ components of $\rho$ and $\ell$ are zero,
Thus the function \eqref{def_Z_rho} becomes
\begin{equation}
	\begin{split}
		{Z_\rho}^\oa{v}_\ob{v}(\tau)
			& =\sum_{\ell\in{\Lambda_\ell}^\oa{v}_\ob{v}}
			\sum_{\ell'\in{\Lambda_{\ell'}}^\oa{v}_\ob{v}}
			\xi^{\oa{v},\ell}_{\ob{v},\ell'} C^{\oa{v},\ell}_{\ob{v},\ell'}
			\\
		& \qquad \times \int d^d\rho^{}_0 d^d\bar{\rho^{}_0}\,
			q^{\bar{(\rho^{}_0+\ell/2)}\bullet(\rho^{}_0+\ell/2)}
			\bar{q}^{\bar{(\rho^{}_0-\ell/2)}\bullet(\rho^{}_0-\ell/2)}
			e^{-2\pi i(\bar{\ell'}\bullet\rho^{}_0
			+\ell'\bullet\bar{\rho^{}_0})},
	\end{split}
	\label{Z_rho0}
\end{equation}
where $d$ is the dimension of $\oa{v^a}=\ob{v^a}=0$ space.
The $\oa{v^a}\ne0$ or $\ob{v^a}\ne0$ components of $\ell$ and $\ell'$
contribute to only the factor
$\xi^{\oa{v},\ell}_{\ob{v},\ell'} C^{\oa{v},\ell}_{\ob{v},\ell'}$
in ${Z_\rho}^\oa{v}_\ob{v}(\tau)$.
Therefore I sum up this factor over $\ell$ and $\ell'$ whose
$\oa{v^a}=\ob{v^a}=0$ components are fixed.
\begin{equation}
	B^{\oa{v},\ell_0}_{\ob{v},\ell'_0}
		=\sum_{\ell\in{\Lambda_\ell}^\oa{v}_\ob{v}(\ell_0)}
		\sum_{\ell'\in{\Lambda_{\ell'}}^\oa{v}_\ob{v}(\ell'_0)}
		\xi^{\oa{v},\ell}_{\ob{v},\ell'} C^{\oa{v},\ell}_{\ob{v},\ell'},
\end{equation}
where $\ell_0$ and $\ell'_0$ are the vectors of $\ell$ and $\ell'$
respectively projected onto $\oa{v^a}=\ob{v^a}=0$ space,
\begin{subequations}
\begin{align}
	{\Lambda_\ell}^\oa{v}_\ob{v}(\ell_0) & =\Set{\ell}
		{\ell\in{\Lambda_\ell}^\oa{v}_\ob{v}, \; \forall a \;
		(\oa{v^a}=\ob{v^a}=0), \; \ell^a=\ell_0^a}, \\
	{\Lambda_{\ell'}}^\oa{v}_\ob{v}(\ell'_0) & =\Set{\ell'}
		{\ell'\in{\Lambda_{\ell'}}^\oa{v}_\ob{v}, \; \forall a \;
		(\oa{v^a}=\ob{v^a}=0), \; \ell^{\prime a}=\ell_0^{\prime a}}.
\end{align}
\end{subequations}
$C^{\oa{v},\ell}_{\ob{v},\ell'}$ must be such constants as
$B^{\oa{v},\ell_0}_{\ob{v},\ell'_0}$ is convergent.
Therefore $\ell$ and $\ell'$ dependence of this constant is necessary for
convergence of ${Z_\rho}^\oa{v}_\ob{v}(\tau)$.
Then the function \eqref{Z_rho0} becomes
\begin{equation}
	{Z_\rho}^\oa{v}_\ob{v}(\tau)
		=\sum_{\ell_0\in{\Lambda_{\ell0}}^\oa{v}_\ob{v}}
		\sum_{\ell'_0\in{\Lambda_{\ell'0}}^\oa{v}_\ob{v}}
		B^{\oa{v},\ell_0}_{\ob{v},\ell'_0}
		\int d^d\rho^{}_0 d^d\bar{\rho^{}_0}\,
		q^{|\rho^{}_0+\ell_0/2|^2} \bar{q}^{|\rho^{}_0-\ell_0/2|^2}
		e^{-2\pi i(\bar{\ell'_0}\cdot\rho^{}_0+\ell'_0\cdot\bar{\rho^{}_0})},
		\label{Z_rho1}
\end{equation}
where ${\Lambda_{\ell0}}^\oa{v}_\ob{v}$ and ${\Lambda_{\ell'0}}^\oa{v}_\ob{v}$
are the lattices of ${\Lambda_{\ell}}^\oa{v}_\ob{v}$ and
${\Lambda_{\ell'0}}^\oa{v}_\ob{v}$ respectively projected onto
$\oa{v^a}=\ob{v^a}=0$ space.
Since $\ell_0$, $\ell'_0$ and $\rho^{}_0$ are the vectors in
$\oa{v^a}=\ob{v^a}=0$ space,
the product of these is $x_0\bullet y_0=x_0\cdot y_0$.
In particular, in case of $\oa{v^a}\ne0$ or $\ob{v^a}\ne0$ for all $a$,
the function \eqref{Z_rho1} reads
\begin{equation}
	{Z_\rho}^\oa{v}_\ob{v}(\tau)=B^{\oa{v},0}_{\ob{v},0},
		\label{Z_rho2}
\end{equation}
because $\rho^{}_0$, $\ell_0$ and $\ell'_0$ do not exist.
Thus I have derived the function ${Z_\rho}^\oa{v}_\ob{v}(\tau)$.

In case of the toroidal compactification ($\alpha=\beta=0$),
${\Lambda_{\ell'0}}^0_0=\Lambda$ because of \eqref{Lambda_lb}.
Therefore, if $B^{0,\ell_0}_{0,\ell'_0}$ do not depend on $\ell'_0$ then
the summation over $\ell'_0$ in the function \eqref{Z_rho1} is
\begin{equation}
	\sum_{\ell'_0\in\Lambda} e^{-2\pi i(\bar{\ell'_0}\cdot\rho^{}_0
		+\ell'_0\cdot\bar{\rho^{}_0})}
		=V_{\Lambda^*} \sum_{\ell''_0\in\Lambda^*}
		\delta(\rho^{}_0-\ell''_0),
\end{equation}
where $\Lambda^*$ is the dual-lattice of $\Lambda$,
and $V_{\Lambda^*}$ is the volume of a unit cell of $\Lambda^*$.
Hence, only $\rho^{}_0$ included in $\Lambda^*$ contribute to the
function ${Z_\rho}^0_0(\tau)$.
Therefore only the invariant states under $h_{\rho;\,\oa{v}}(\ob{v},\ell')$
contribute to the partition function,
because if $\rho^{}_0\in\Lambda^*$ then the eigenvalue of
$h_{\rho;\,\oa{v}}(\ob{v},\ell')$ is
$e^{-2\pi i(\bar{\ell'_0}\cdot\rho^{}_0+\ell'_0\cdot\bar{\rho^{}_0})}=1$.
In the orbifold compactification, however, we must also consider states which
is not invariant under $h_{\rho;\,\oa{v}}(\ob{v},\ell')$.
The factor
$e^{-2\pi i(\bar{\ell'_0}\cdot\rho^{}_0+\ell'_0\cdot\bar{\rho^{}_0})}$ is
necessary for modular invariance of the partition function (see the next
section).

Second, I calculate the function ${Z_N}^\oa{v}_\ob{v}(\tau)$.
The function ${Z_N}^\oa{v}_\ob{v}(\tau)$ can be derived
by calculating the trace in this function \eqref{def_Z_N} by attention to
$N_0^{\prime a}=\tilde{N}_0^a=0$,
\begin{align}
	{Z_N}^\oa{v}_\ob{v}(\tau)
		& =(q\bar{q})^{\oa{v}\cdot(1-\oa{v})/2-3/12}
		\prod_{a=1}^3 \prod_{n=1}^\infty \left\{
		\delta_{\oa{v^a},0} \absolute{\kakko{1-q^n e^{2\pi i\ob{v^a}}}^{-1}
		\kakko{1-q^n e^{-2\pi i\ob{v^a}}}^{-1}}^2 \right. \notag \\
	& \qquad \left. +(1-\delta_{\oa{v^a},0})
		\absolute{\kakko{1-q^{n-\oa{v^a}}e^{2\pi i\ob{v^a}}}^{-1}
		\kakko{1-q^{n+\oa{v^a}-1}e^{-2\pi i\ob{v^a}}}^{-1}}^2 \right\}.
		\label{Z_N}
\end{align}
Thus I have derived the function ${Z_N}^\oa{v}_\ob{v}(\tau)$.

I have shown that the partition function $Z^\oa{v}_\ob{v}(\tau)$ is given by
product of the functions \eqref{Z_rho1} and \eqref{Z_N}.
Finally, I conclude that the complete total partition function $Z(\tau)$
is given by summation of the function $Z^\oa{v}_\ob{v}(\tau)$.

\section{Modular invariance conditions}

In this section, I derive modular invariance conditions of the total partition
function $Z(\tau)$.
The modular transformation is composed of the following two transformations:
the one is $\tau\to\tau+1$ called T-transformation, the other is
$\tau\to-1/\tau$ called S-transformation.

\subsection{T-transformation}

First, I calculate the T-transformation of the functions
${Z_\rho}^\oa{v}_\ob{v}(\tau)$, ${Z_N}^\oa{v}_\ob{v}(\tau)$ and
$Z^\oa{v}_\ob{v}(\tau)$.
In case of $\oa{v^a}=\ob{v^a}=0$ for some $a$, the T-transformation of the
function \eqref{Z_rho1} becomes
\begin{align}
	& {Z_\rho}^\oa{v}_\ob{v}(\tau+1) \notag \\
	& =\sum_{\ell_0\in{\Lambda_{\ell0}}^\oa{v}_\ob{v}}
		\sum_{\ell'_0\in{\Lambda_{\ell'0}}^\oa{v}_\ob{v}}
		B^{\oa{v},\ell_0}_{\ob{v},\ell'_0}
		\int d^d\rho^{}_0 d^d\bar{\rho^{}_0}\,
		e^{2\pi i(\tau+1)|\rho^{}_0+\ell_0/2|^2}
		e^{-2\pi i(\bar{\tau}+1)|\rho^{}_0-\ell_0/2|^2}
		e^{-2\pi i(\bar{\ell'_0}\cdot\rho^{}_0
		+\ell'_0\cdot\bar{\rho^{}_0})} \notag \\
	& =\sum_{\ell_0\in{\Lambda_{\ell0}}^\oa{v}_\ob{v}}
		\sum_{\ell'_0\in{\Lambda_{\ell'0}}^\oa{v}_\ob{v}}
		B^{\oa{v},\ell_0}_{\ob{v},\ell'_0}
		\int d^d\rho^{}_0 d^d\bar{\rho^{}_0}\,
		q^{|\rho^{}_0+\ell_0/2|^2} \bar{q}^{|\rho^{}_0-\ell_0/2|^2}
		e^{-2\pi i\{\bar{(\ell'_0-\ell_0)}\cdot\rho^{}_0
		+(\ell'_0-\ell_0)\cdot\bar{\rho^{}_0}\}}.
\end{align}
If $\xi^{\oa{v},\ell}_{\ob{v},\ell'}=1$ then $\ell^{\prime a}-\ell^a=0$
at $\oa{v^a}\ne0,\,\oba{v^a}=0$ because of \eqref{def_xi}.
Therefore, if $\xi^{\oa{v},\ell}_{\ob{v},\ell'}=1$ then
$\ell'-\ell\in{\Lambda_{\ell'}}^\oa{v}_\oba{v}$ owing to \eqref{Lambda_lb}.
Hence, let $\ell''_0$ be $\ell'_0-\ell_0$,
then $\ell''_0\in{\Lambda_{\ell'0}}^\oa{v}_\oba{v}$.
In addition,
${\Lambda_{\ell0}}^\oa{v}_\ob{v}={\Lambda_{\ell0}}^\oa{v}_\oba{v}$ because of
\eqref{Lambda_la}.
Thus this T-transformation becomes
\begin{equation}
	{Z_\rho}^\oa{v}_\ob{v}(\tau+1)
		=\sum_{\ell_0\in{\Lambda_{\ell0}}^\oa{v}_\oba{v}}
		\sum_{\ell''_0\in{\Lambda_{\ell'0}}^\oa{v}_\oba{v}}
		B^{\oa{v},\ell_0}_{\ob{v},\ell''_0+\ell_0}
		\int d^d\rho^{}_0 d^d\bar{\rho^{}_0}\,
		q^{|\rho^{}_0+\ell_0/2|^2} \bar{q}^{|\rho^{}_0-\ell_0/2|^2}
		e^{-2\pi i(\bar{\ell''_0}\cdot\rho^{}_0
		+\ell''_0\cdot\bar{\rho^{}_0})}.
\end{equation}
I assume the condition $B^{\oa{v},\ell_0}_{\ob{v},\ell''_0+\ell_0}
=F_T B^{\oa{v},\ell_0}_{\oba{v},\ell''_0}$ for all $\ell_0$ and $\ell''_0$,
where $F_T$ is a constant.
Then this T-transformation is
${Z_\rho}^\oa{v}_\ob{v}(\tau+1)=F_T{Z_\rho}^\oa{v}_\oba{v}(\tau)$.
This constant is $F_T=B^{\oa{v},0}_{\ob{v},0}/B^{\oa{v},0}_{\oba{v},0}$,
because if $\ell_0=\ell''_0=0$ then
$B^{\oa{v},0}_{\ob{v},0}=F_T B^{\oa{v},0}_{\oba{v},0}$.
Then this T-transformation is given by
\begin{equation}
	{Z_\rho}^\oa{v}_\ob{v}(\tau+1)
		=\frac{B^{\oa{v},0}_{\ob{v},0}}{B^{\oa{v},0}_{\oba{v},0}}
		{Z_\rho}^\oa{v}_\oba{v}(\tau).
		\label{Z_rho_T}
\end{equation}
The condition $B^{\oa{v},\ell_0}_{\ob{v},\ell''_0+\ell_0}
=F_T B^{\oa{v},\ell_0}_{\oba{v},\ell''_0}$ becomes
\begin{equation}
	\frac{B^{\oa{v},\ell_0}_{\ob{v},\ell''_0+\ell_0}}{B^{\oa{v},0}_{\ob{v},0}}
		=\frac{B^{\oa{v},\ell_0}_{\oba{v},\ell''_0}}{B^{\oa{v},0}_{\oba{v},0}}
		. \label{condition_Tl}
\end{equation}
In case of $\oa{v^a}\ne0$ or $\ob{v^a}\ne0$ for all $a$,
the function \eqref{Z_rho2} are transformed equally as \eqref{Z_rho_T}.
Thus I have derived the T-transformation of the function
${Z_\rho}^\oa{v}_\ob{v}(\tau)$.

The T-transformation of the function \eqref{Z_N} can be easily
calculated,
\begin{equation}
	{Z_N}^\oa{v}_\ob{v}(\tau+1)={Z_N}^\oa{v}_\oba{v}(\tau). \label{Z_N_T}
\end{equation}
Thus I have derived the T-transformation of the function
${Z_N}^\oa{v}_\ob{v}(\tau)$.

By using the transformations \eqref{Z_rho_T} and \eqref{Z_N_T}, the
T-transformation of the partition function $Z^\oa{v}_\ob{v}(\tau)$ is given by
\begin{equation}
	Z^\oa{v}_\ob{v}(\tau+1)
		=\frac{B^{\oa{v},0}_{\ob{v},0}}{B^{\oa{v},0}_{\oba{v},0}}
		Z^\oa{v}_\oba{v}(\tau).
		\label{Z_T}
\end{equation}
Thus I have derived the T-transformation of the partition function
$Z^\oa{v}_\ob{v}(\tau)$.

\subsection{S-transformation}

Second, I calculate the S-transformations of the function
${Z_\rho}^\oa{v}_\ob{v}(\tau)$, ${Z_N}^\oa{v}_\ob{v}(\tau)$ and
$Z^\oa{v}_\ob{v}(\tau)$.
In case of $\oa{v^a}=\ob{v^a}=0$ for some $a$, by integrating over
$\rho^{}_0$, the function \eqref{Z_rho1} becomes
\begin{equation}
	\begin{split}
		{Z_\rho}^\oa{v}_\ob{v}(\tau)
			& =\frac{1}{(2\tau_2)^d}
			\sum_{\ell_0\in{\Lambda_{\ell0}}^\oa{v}_\ob{v}}
			\sum_{\ell'_0\in{\Lambda_{\ell'0}}^\oa{v}_\ob{v}}
			{B_\phi}^{\oa{v},\ell_0}_{\ob{v},\ell'_0} \\
		& \qquad \times \exp\Kakko{-\frac{\pi}{\tau_2}(|\tau|^2|\ell_0|^2
			-\tau_1\bar{\ell_0}\cdot\ell'_0-\tau_1\ell_0\cdot\bar{\ell'_0}
			+|\ell'_0|^2)},
	\end{split}
	\label{Z_rho3}
\end{equation}
where $\tau=\tau_1+i\tau_2$.
The S-transformations of $\tau_1$ and $\tau_2$ are $\tau_1\to-\tau_1/|\tau|^2$
and $\tau_2\to\tau_2/|\tau|^2$.
Therefore the S-transformation of the function \eqref{Z_rho3} becomes
\begin{equation}
	\begin{split}
		{Z_\rho}^\oa{v}_\ob{v}(-1/\tau)
			& =\kakko{\frac{|\tau|^2}{2\tau_2}}^d
			\sum_{\ell_0\in{\Lambda_{\ell0}}^\oa{v}_\ob{v}}
			\sum_{\ell''_0\in{\Lambda_{\ell'0}}^\oa{v}_\ob{v}}
			B^{\oa{v},\ell_0}_{\ob{v},\ell'_0} \\
		& \qquad \times \exp\Kakko{-\frac{\pi}{\tau_2}\kakko{|\ell_0|^2
			+\tau_1\bar{\ell_0}\cdot\ell'_0+\tau_1\ell_0\cdot\bar{\ell'_0}
			+|\tau|^2|\ell'_0|^2}}.
	\end{split}
\end{equation}
I change $\ell_0$ and $\ell'_0$ to $-\ell'_0$ and $\ell_0$, respectively.
Then $\ell_0\in{\Lambda_{\ell'0}}^\oa{v}_\ob{v}
={\Lambda_{\ell0}}^\ob{v}_\oma{v}$ and
$\ell'_0\in{\Lambda_{\ell0}}^\oa{v}_\ob{v}={\Lambda_{\ell'0}}^\ob{v}_\oma{v}$
because of \eqref{Lambda_la} and \eqref{Lambda_lb}.
Thus this S-transformation becomes
\begin{equation}
	\begin{split}
		{Z_\rho}^\oa{v}_\ob{v}(-1/\tau)
			& =\kakko{\frac{|\tau|^2}{2\tau_2}}^d
			\sum_{\ell_0\in{\Lambda_{\ell0}}^\ob{v}_\oma{v}}
			\sum_{\ell'_0\in{\Lambda_{\ell'0}}^\ob{v}_\oma{v}}
			B^{\oa{v},-\ell'_0}_{\ob{v},\ell_0} \\
		& \qquad \times \exp\Kakko{-\frac{\pi}{\tau_2}
			\kakko{|\tau|^2|\ell_0|^2-\tau_1\bar{\ell_0}\cdot\ell'_0
			-\tau_1\ell_0\cdot\bar{\ell'_0}+|\ell'_0|^2}}.
	\end{split}
\end{equation}
I assume the condition $B^{\oa{v},-\ell'_0}_{\ob{v},\ell_0}
=F_S B^{\ob{v},\ell_0}_{\oma{v},\ell'_0}$ for all $\ell_0$ and $\ell'_0$,
where $F_S$ is a constant.
Then this S-transformation is
${Z_\rho}^\oa{v}_\ob{v}(-1/\tau)=F_S|\tau|^{2d}{Z_\rho}^\ob{v}_\oma{v}(\tau)$.
This constant is $F_S=B^{\oa{v},0}_{\ob{v},0}/B^{\ob{v},0}_{\oma{v},0}$,
because if $\ell_0=\ell'_0=0$ then
$B^{\oa{v},0}_{\ob{v},0}=F_S B^{\ob{v},0}_{\oma{v},0}$.
Then this S-transformation is given by
\begin{equation}
	{Z_\rho}^\oa{v}_\ob{v}(-1/\tau)
		=\frac{B^{\oa{v},0}_{\ob{v},0}}{B^{\ob{v},0}_{\oma{v},0}} |\tau|^{2d}
		{Z_\rho}^\ob{v}_\oma{v}(\tau).
		\label{Z_rho_S}
\end{equation}
The condition $B^{\oa{v},-\ell'_0}_{\ob{v},\ell_0}
=F_S B^{\ob{v},\ell_0}_{\oma{v},\ell'_0}$ becomes
\begin{equation}
	\frac{B^{\oa{v},-\ell'_0}_{\ob{v},\ell_0}}{B^{\oa{v},0}_{\ob{v},0}}
		=\frac{B^{\ob{v},\ell_0}_{\oma{v},\ell'_0}}{B^{\ob{v},0}_{\oma{v},0}}.
		\label{condition_Sl}
\end{equation}
In case of $\oa{v^a}\ne0$ or $\ob{v^a}\ne0$ for all $a$,
the function \eqref{Z_rho2} are transformed equally as \eqref{Z_rho_S}
because of $d=0$.
Thus I have derived the S-transformation of the function
${Z_\rho}^\oa{v}_\ob{v}(\tau)$.

Now I rewrite the function \eqref{Z_N} with the Dedekind eta function
$\eta(\tau)$ and the theta function $\vartheta(\nu,\tau)$.
The Dedekind eta function and the theta function are defined as
\begin{subequations}
\begin{align}
	\eta(\tau) & =q^{1/24}\prod_{n=1}^\infty(1-q^n), \\
	\vartheta(\nu,\tau) & =\prod_{n=1}^\infty
		(1-q^n)(1+e^{2\pi i\nu}q^{n-1/2})(1+e^{-2\pi i\nu}q^{n-1/2}).
\end{align}
\end{subequations}
Then the function ${Z_N}^\oa{v}_\ob{v}(\tau)$ becomes
\begin{align}
	{Z_N}^\oa{v}_\ob{v}(\tau)
		=\prod_{a=1}^3 \left\{\frac{\delta_{\oa{v^a},0}
		\delta_{\ob{v^a},0}}{|[\eta(\tau)]^2|^2} \right.
		& +(q\bar{q})^{-(\oa{v^a}-1/2)^2/2}
		\kakko{\delta_{\oa{v^a},0}(1-\delta_{\ob{v^a},0})
		4\sin^2\pi\ob{v^a}+(1-\delta_{\oa{v^a},0})} \notag \\
	& \qquad \left. \times \absolute{\frac{\eta(\tau)}
		{\vartheta((-\oa{v^a}+1/2)\tau+\ob{v^a}+1/2,\tau)}}^2 \right\}.
		\label{Z_N_theta}
\end{align}
The S-transformation of the function \eqref{Z_N_theta} can be
calculated by using the formulas $\eta(-1/\tau)=(-i\tau)^{1/2}\eta(\tau)$,
$\vartheta(-\nu,\tau)=\vartheta(\nu,\tau)$ and
$\vartheta(\nu/\tau,-1/\tau)
=(-i\tau)^{1/2}e^{\pi i\nu^2/\tau}\vartheta(\nu,\tau)$,
\begin{align}
	{Z_N}^\oa{v}_\ob{v}(-1/\tau)
		& =\prod_{a=1}^3
		\left\{\frac{\delta_{\oa{v^a},0} \delta_{\ob{v^a},0}}{|\tau|^2}
		+\delta_{\oa{v^a},0}(1-\delta_{\ob{v^a},0})
		4\sin^2\pi\ob{v^a} \right. \notag \\
	& \qquad \left. +\frac{(1-\delta_{\oa{v^a},0})\delta_{\ob{v^a},0}}
		{4\sin^2\pi\oa{v^a}}
		+(1-\delta_{\oa{v^a},0})(1-\delta_{\ob{v^a},0}) \right\}
		{Z_N}^\ob{v}_\oma{v}(\tau) \notag \\
	& =\frac{F^\oa{v}_\ob{v}}{F^\ob{v}_\oma{v}} \frac{1}{|\tau|^{2d}}
		{Z_N}^\ob{v}_\oma{v}(\tau),
		\label{Z_N_S}
\end{align}
where
\begin{equation}
	F^\oa{v}_\ob{v}=\prod_{a=1}^3
		\Kakko{\delta_{\oa{v^a},0}(1-\delta_{\ob{v^a},0}) 4\sin^2\pi\ob{v^a}
		+\delta_{\oa{v^a},0} \delta_{\ob{v^a},0}
		+(1-\delta_{\oa{v^a},0})}.
\end{equation}
Thus I have derived the S-transformation of the function
${Z_N}^\oa{v}_\ob{v}(\tau)$.

By using the transformations \eqref{Z_rho_S} and \eqref{Z_N_S},
the S-transformation of the partition function $Z^\oa{v}_\ob{v}(\tau)$ is
given by
\begin{equation}
	Z^\oa{v}_\ob{v}(-1/\tau)
		=\frac{B^{\oa{v},0}_{\ob{v},0}}{B^{\ob{v},0}_{\oma{v},0}}
		\frac{F^\oa{v}_\ob{v}}{F^\ob{v}_\oma{v}}
		Z^\ob{v}_\oma{v}(\tau).
		\label{Z_S}
\end{equation}
Thus I have derived the S-transformation of the partition function
$Z^\oa{v}_\ob{v}(\tau)$.

\subsection{Modular invariance conditions}

Finally, I derive the modular invariance conditions of the total partition
function \eqref{def_Z}.
By using the T-transformation \eqref{Z_T},
the T-transformation of the total partition function $Z(\tau)$ becomes
\begin{equation}
	Z(\tau+1)=\frac{1}{N} \sum_{\alpha=0}^{N-1} \sum_{\beta=0}^{N-1}
		\frac{B^{\oa{v},0}_{\ob{v},0}}{B^{\oa{v},0}_{\oba{v},0}}
		Z^\oa{v}_\oba{v}(\tau).
\end{equation}
I find that the condition $B^{\oa{v},0}_{\ob{v},0}=B^{\oa{v},0}_{\oba{v},0}$
implies T-invariance of $Z(\tau)$: $Z(\tau+1)=Z(\tau)$.
By using this condition, the condition \eqref{condition_Tl} becomes
$B^{\oa{v},\ell_0}_{\ob{v},\ell''_0+\ell_0}
=B^{\oa{v},\ell_0}_{\oba{v},\ell''_0}$.
In fact, the condition $B^{\oa{v},\ell_0}_{\ob{v},\ell''_0+\ell_0}
=B^{\oa{v},\ell_0}_{\oba{v},\ell''_0}$ includes the condition
$B^{\oa{v},0}_{\ob{v},0}=B^{\oa{v},0}_{\oba{v},0}$ itself as a special case.

Similarly, by using the S-transformation \eqref{Z_S},
the S-transformation of the total partition function $Z(\tau)$ becomes
\begin{equation}
	Z(-1/\tau)=\frac{1}{N} \sum_{\alpha=0}^{N-1} \sum_{\beta=0}^{N-1}
		\frac{B^{\oa{v},0}_{\ob{v},0}}{B^{\ob{v},0}_{\oma{v},0}}
		\frac{F^\oa{v}_\ob{v}}{F^\ob{v}_\oma{v}} Z^\ob{v}_\oma{v}(\tau).
\end{equation}
I find that the condition $B^{\oa{v},0}_{\ob{v},0}F^\oa{v}_\ob{v}
=B^{\ob{v},0}_{\oma{v},0}F^\ob{v}_\oma{v}$
implies S-invariance of $Z(\tau)$: $Z(-1/\tau)=Z(\tau)$.
By using this condition, the condition \eqref{condition_Sl} becomes
$B^{\oa{v},-\ell'_0}_{\ob{v},\ell_0}F^\oa{v}_\ob{v}
=B^{\ob{v},\ell_0}_{\oma{v},\ell'_0}F^\ob{v}_\oma{v}$.
In fact, the condition $B^{\oa{v},-\ell'_0}_{\ob{v},\ell_0}F^\oa{v}_\ob{v}
=B^{\ob{v},\ell_0}_{\oma{v},\ell'_0}F^\ob{v}_\oma{v}$ includes the condition
$B^{\oa{v},0}_{\ob{v},0}F^\oa{v}_\ob{v}
=B^{\ob{v},0}_{\oma{v},0}F^\ob{v}_\oma{v}$ itself as a special case.

Thus I have derived sufficient conditions for the modular invariance of the
total partition function are
\begin{equation}
	B^{\oa{v},\ell_0}_{\ob{v},\ell''_0+\ell_0}
		=B^{\oa{v},\ell_0}_{\oba{v},\ell''_0}, \quad
		B^{\oa{v},-\ell'_0}_{\ob{v},\ell_0}F^\oa{v}_\ob{v}
		=B^{\ob{v},\ell_0}_{\oma{v},\ell'_0}F^\ob{v}_\oma{v}.
		\label{inv_condition}
\end{equation}

\section{Summary and remarks}

I have constructed the complete 1-loop partition function of a bosonic closed
string on orbifolds.
In particular, I have paid attention to the following points:
the $\ell$ and $\ell'$ dependence of the constant
$C^{\oa{v},\ell}_{\ob{v},\ell'}$,
the derivation of the eigenstates and eigenvalues of the operator
$h_{\rho;\,\oa{v}}(\ob{v},\ell')$,
and existence of the fixed points which satisfy both conditions
\eqref{fixed_point_a} and \eqref{fixed_point_b}.
Furthermore, I have derived sufficient conditions \eqref{inv_condition} for
the modular invariance of the total partition function.

In this paper, I particularly discussed the six internal coordinates in the
heterotic string.
If this argument is adapted to sixteen left-moving bosons (corresponding to
sixteen internal coordinates) in the heterotic strings,
it is expected to derive higher level current algebras naturally
\cite{bosonic}.
The higher level current algebras are necessary for grand unified models in
the 4-dimensional heterotic string \cite{higher_level}.
I think that the $\ell$ and $\ell'$ dependence of the constant
$C^{\oa{v},\ell}_{\ob{v},\ell'}$ is necessary for modular invariance of the
left-moving bosons on orbifolds.

\section*{Acknowledgment}

I would like to thank K.I. Kondo for helpful advice.


\end{document}